\documentclass[9pt,twocolumn,twoside]{osajnl}

\journal{ol} 

\setboolean{shortarticle}{true}
\usepackage{bm}           
\usepackage{braket}       
\usepackage{color}        
\usepackage{siunitx}      
\usepackage{float}

\title{Generation of Photon Pairs by Stimulated Emission in Ring Resonators}

\author[1,*]{Milica Banic}
\author[2]{Marco Liscidini}
\author[1]{J. E. Sipe}

\affil[1]{Department of Physics, University of Toronto, Toronto, ON, M5S 1A7, Canada}
\affil[2]{Department of Physics, University of Pavia, I-27100 Pavia, Italy}

\affil[*]{Corresponding author: mbanic@physics.utoronto.ca}

\begin{abstract}
Third-order parametric down-conversion (TOPDC) describes a class of nonlinear interactions in which a pump photon is converted into a photon triplet. This process can occur spontaneously, or it can be stimulated by seeding fields. 
Here we show that stimulated TOPDC (StTOPDC) can be exploited for the generation of quantum correlated photon pairs. {We model StTOPDC in a microring resonator, predicting observable pair generation rates in a microring engineered for third-harmonic generation, and we examine the peculiar features of this approach when compared with second-order spontaneous parametric down-conversion and spontaneous four-wave mixing. We conclude that if the experimental difficulties associated with implementing StTOPDC can be overcome, it may soon be possible to demonstrate this process in resonant integrated devices.}
\end{abstract}

\setboolean{displaycopyright}{true}

\begin{document}

\maketitle

\maketitle

Nonlinear parametric processes are arguably the most common phenomena used for the generation of nonclassical states of light. Practical implementations typically rely on two such processes: spontaneous parametric down-conversion (SPDC), a second-order process in which a pump photon is converted to a photon pair, and  spontaneous four-wave mixing (SFWM), a third-order process in which the annihilation of two pump photons again results in the generation of a photon pair. Both SPDC and SFWM have been demonstrated in a plethora of systems, ranging from bulk optics \cite{kwiat_new_1995} to integrated photonics \cite{grassani_micrometer-scale_2015}. 

For a third-order nonlinearity, one can envision an alternative approach to generating non-classical light based on third-order parametric down conversion (TOPDC), in which one pump photon is  converted to a photon triplet \cite{Chekhova_pra_2005}. Spontaneous TOPDC (SpTOPDC) is in fact the third-order analogue of SPDC, but with the important difference that photons are generated in triplets instead of pairs. Thus, in general the generated state is non-Gaussian. This feature has inspired considerable interest in SpTOPDC in recent years, because non-Gaussian states can be very powerful resources in quantum computation, but are difficult to generate on-demand \cite{hamel_np_2014,okoth_pra_2019,Walschaers_arXiv_21, threephoton_1997}. However, SpTOPDC is a very inefficient process, because it relies on a third-order nonlinear response with a generation rate that scales linearly with the pump intensity. A few strategies have been proposed to improve the process efficiency \cite{corona_pra_2011,Corona_ol_2011,dot_pra_2012}, but practical generation rates have not yet been demonstrated.

Here our focus is instead on stimulated TOPDC (StTOPDC). In this scenario, photons are still emitted in triplets, but the down-conversion of the pump photon is stimulated by the presence of additional seed fields which, unlike the pump field, do not provide any energy but simply stimulate the process \cite{Blay_StTOPDC}. Recently, it has been shown that StTOPDC can be exploited to characterize the properties of the quantum light that would be generated by SpTOPDC in the absence of any seed field \cite{dominguez-serna_third-order_2020, liscidini_stimulated_2013}. In contrast to that, here we are interested in StTOPDC as a scheme to generate quantum correlated photon pairs.

{Whereas earlier studies of TOPDC have dealt only with non-resonant structures, here we investigate the use of microring resonators to enhance and control StTOPDC \cite{Corona_ol_2011,okoth_pra_2019,corona_pra_2011}; addressing this resonant configuration is essential due to the qualitative differences that generally arise between nonlinear processes in resonant and non-resonant systems \cite{how_does_it_scale}.} We calculate the StTOPDC
photon pair generation rate for a composite AlN-Si$_3$N$_4$ structure compatible with the current fabrication technology, and characterize the biphoton state arising from StTOPDC in this structure by calculating the joint spectral intensity (JSI).

We begin by considering the generic Hamiltonian describing the generation of photon triplets via TOPDC, in which a pump photon in mode $P$ is down-converted into three photons, one in each of the modes $G_1$, $G_2$, and $G_3$:
\begin{align}
    H_{TOPDC} = - \hbar \sum_{G_1,G_2,G_3} \Lambda_{G_1 G_2 G_3 P}
    a^{\dagger}_{G_1} a^{\dagger}_{G_2} a^{\dagger}_{G_3} a_P + H.c.\label{eq: spon_TG}, 
\end{align}
where $a^{\dagger}_i$ is the creation operator of a photon in the $i$-th mode and  $\Lambda_{G_1 G_2 G_3 P}$ is the nonlinear coupling rate, which depends on the structure under consideration. 

Most studies have focused on fully degenerate ($G_1 = G_2 = G_3$)  or fully non-degenerate ($G_1 \neq G_2 \neq G_3$) SpTOPDC, and the tripartite entanglement characterizing the generated triplet \cite{Moebius:16}.  However, another scenario is that in which the down-conversion is stimulated by a seed field  \cite{okoth_pra_2019}, with the other two photons being quantum correlated and generated in the modes $G_1$ and $G_2$. For that process it is useful to rewrite \eqref{eq: spon_TG} as
\begin{align}
    H_{TOPDC} = - 3\hbar \sum_{G_1,G_2\neq S} \Lambda_{G_1 G_2 S P}
    a^{\dagger}_{G_1} a^{\dagger}_{G_2} a^{\dagger}_{S} a_P + H.c. \label{eq: stim_TG},
\end{align}
where we highlight the creation operator associated with the mode $S$ that would be seeded in the stimulated process, and we restrict the discussion to configurations where $G_1, G_2 \neq S$. If one traces over all the photons generated in the mode $S$, StTOPDC can be thought as an effective 
photon pair generation process.

\begin{figure}
    \centering
    \includegraphics[width=0.5\textwidth]{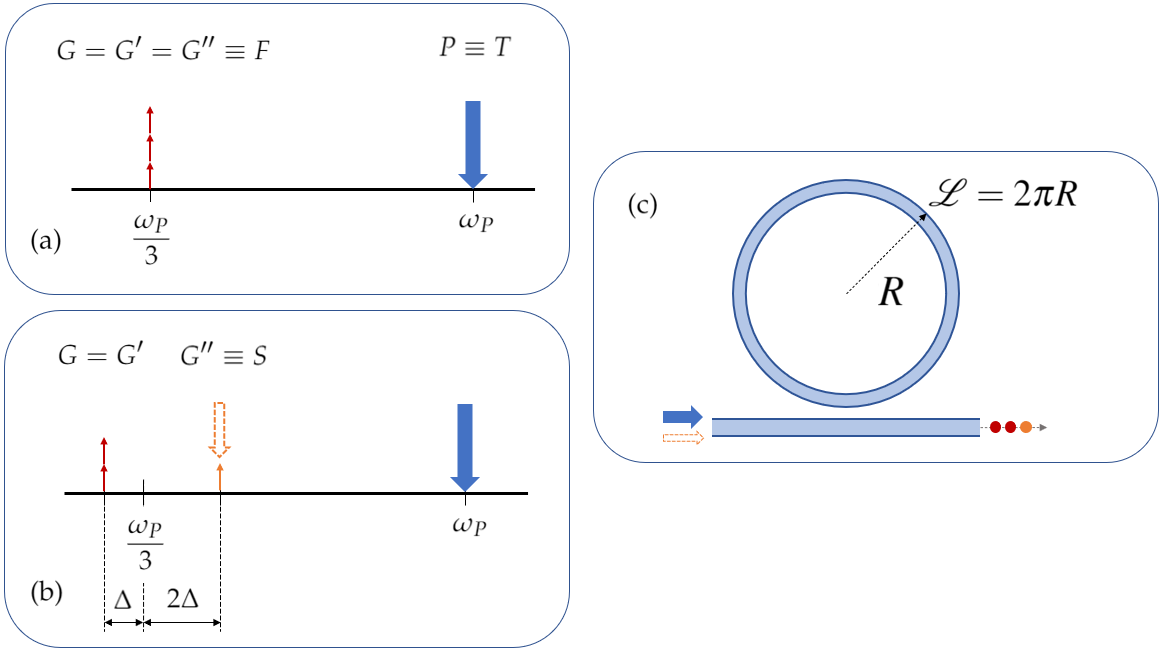}
    \caption{Schematics of the mode configurations for 
    (a) fully degenerate (b) non-degenerate TOPDC
    , and (c) a sketch of the ring resonator 
    in the non-degenerate configuration. 
    Arrows pointing up represent generated modes; arrows pointing down represent input modes. The dotted arrow in (b) represents the possibility of injecting a seed field in mode $S$.}
    \label{fig:configuration}
\end{figure}

{One might assume at first glance that StTOPDC, when driven by classical pump and seed fields, is equivalent to dual-pump SFWM. However, although the Hamiltonians describing the two processes in this scenario are formally similar, it is important to stress that the seed field cannot be understood as a second pump; indeed, the seed does not provide any energy in the generation of the photon pairs, and is in fact amplified by StTOPDC. This distinction between StTOPDC and SFWM is due to the fact that the former stems from a `higher order' spontaneous process.}

This has practical consequences. First, because the generation of pairs is seeded, the properties of the pairs generated by StTOPDC are sensitive to those of the seed field. Second, energy conservation leads to an unusual frequency configuration of the four modes involved in the process.
%
%
{We note that unlike in SFWM, here the pump and seed fields can be asymmetrically located with respect to the generated modes (as in the case sketched in Fig. \ref{fig:configuration}b), with the pump field being the farthest. In this case, both input fields could be filtered using a low-pass filter, instead of notch or bandpass filters, which are typically required in SFWM \cite{grassani_micrometer-scale_2015}. In addition, input power can be distributed unevenly between the two inputs, without affecting the efficiency. Thus, for some implementations this frequency configuration may lead to easier filtering stages, particularly in integrated platforms.}
%
%
{The flexibility in the pump and seed frequencies also leads to other practical advantages: For example, it gives more freedom in dispersion engineering to achieve phase matching. In addition, one could use this degree of freedom, along with the ability to redistribute the input powers, to limit the effect of parasitic processes that can arise when photon pairs are generated via SFWM \cite{zhang_nat_comm_2021}; to suppress such processes in SFWM, one must rely on the use of more complex structures \cite{linearly_uncoupled}.}

With these interesting features in mind, we calculate the generation rates for TOPDC. Unlike earlier studies, 
here we focus on an integrated ring resonator structure as sketched in Fig. \ref{fig:configuration}c
\cite{Corona_ol_2011,okoth_pra_2019,corona_pra_2011}. This enables us to take advantage of the field enhancement associated with the spatial and temporal light confinement, and to exploit the presence of a comb of resonances which favours the scenario in which pump, seed, and generated photons are on resonance at well-separated frequencies. We assume that the effect of the nonlinearity can be restricted to the ring, where the field intensities are the largest, so the operators $a$ and $a^{\dagger}$ in (\ref{eq: spon_TG}) and (\ref{eq: stim_TG}) refer to the ring modes.

We begin by using Fermi's Golden Rule to make a first-order calculation of the rates for SpTOPDC and StTOPDC. For simplicity, we consider pump fields and, when present, seed fields in the CW limit. For the spontaneous process governed by Eq. (\ref{eq: spon_TG}), we first consider the generation of triplets in a single fundamental resonance from a third harmonic pump field (i.e. $P \equiv T$ and $G_1=G_2=G_3 \equiv F$, see Fig.\ref{fig:configuration}a). In reality, there will be multiple energy-conserving processes that can generate triplets in sets of different resonances, but we defer the inclusion of these processes to later work. We adopt an interaction picture such that $H_{NL}^{I}(t) = e^{iH_0 t/\hbar} H_{NL} e^{-iH_0 t/\hbar}$, where $H_0$ captures the linear dynamics of the ring-channel system and $H_{NL}$ is the nonlinear Hamiltonian \cite{Yang2008}. Then for this degenerate process we have
\begin{align}
    H_{\text{deg}}^{I}(t) =& 
    \nonumber - \hbar \Lambda_{F F F T} \mathcal{L}^2 \frac{1}{4\pi^2} \text{sinc}\left( \frac{\Delta \kappa\mathcal{L}}{2}\right)  \int dk_1 ... dk_4 F_{F+}^*({k_1})\\
    \nonumber
    &\times F_{F+}^*({k_2})F_{F+}^*({k_3})F_{T-}({k_4}) e^{-i(\omega_{k_4T} - \omega_{k_3F} - \omega_{k_2F} - \omega_{k_1F})t} \\
    &\times a^{\dagger}_F(k_1) a^{\dagger}_F(k_2) a^{\dagger}_F(k_3) \alpha_T(k_4) + H.c.,
    \label{eq:H_NL_spontTG}
\end{align}
where $a^{\dagger}_F(k)$ is the creation operator of a photon in the mode $F$ with wavevector $k$,  and we treat the pump classically by taking $a_T(k_4) \rightarrow \alpha_T(k_4)$, with $\alpha_T(k_4)$ the complex amplitude of the pump field in the channel. Here $\mathcal{L}$ is the ring circumference, and $\Delta \kappa = \kappa_T - 3 \kappa_F$, where $\kappa_J$ is the wavenumber in the ring corresponding to the resonant frequency $\omega_J$. If the channel waveguide and the ring are identical, $\kappa_J \cong K_J$, where $K_J$ is the resonant wavenumber in the channel waveguide. We use $ \omega_{kJ}=\omega_J +v_J(k-K_J)$, where $v_J$ is the group velocity at $\omega_J$. We neglect the effects of group velocity dispersion across each resonance \footnotemark[1], but we take them into account between resonances. 
\footnotetext[1]{{This approximation is valid if the group velocity dispersion $\beta_2$ is small enough; for a 1 GHz resonance linewidth, we require $\beta_2 < 10^{-20}$ s$^2$/m. By numerical simulations, we find $\beta_2 < 10^{-22}$ s$^2$/m for the relevant modes in the sample system considered below.}}
We also introduce complex field enhancement factors
\begin{align}
    F_{J\pm}(k) = \frac{1}{\sqrt{\mathcal{L}}} \left( \frac{\gamma^*_J}{v_J(K_{J} - k) \pm i\overline{\Gamma}_J }\right),
    \label{eq:Lorentzian}
\end{align}
where $\gamma_J$ is the ring-channel coupling constant, and the linewidth of resonance $J$ is set by $\overline{\Gamma}_J =\omega_J/{2Q_J}$, where $Q_J$ is the loaded quality factor \cite{vernon_spontaneous_2015}. Here and in \eqref{eq:H_NL_spontTG}, $\pm$ indicates incoming (+) or outgoing (-) fields \cite{liscidini_pra_2012}. Finally,
$\Lambda_{F F F T}$ is the nonlinear coupling rate 
\begin{align}
    \Lambda_{F F F T} =& 
    \frac{\hbar \sqrt{\omega_{T}\omega_{F}^3}} {4\epsilon_0 c^2} \sqrt{\frac{{v}_F^3 {v}_T}{\overline{n}_F^3 \overline{n}_T}}\frac{\overline{\chi}_3}{\mathcal{L} A_{eff}}, \label{eq:lambda}
\end{align}
%
%
where $\overline{\chi}_3$ is a characteristic value of $\chi_3$ in the ring, and $\overline{n}_T$ and $\overline{n}_F$ are respectively characteristic values of the refractive indices at $\omega_T$ and $\omega_F$; finally, $A_{eff}$ is an effective area determined by the mode overlap in the ring \cite{how_does_it_scale}.

Assuming a CW pump at $\omega_T$, the rate of generated triplets appearing at the system's output is 
\begin{align}
    R^\text{spon}_{F F F} = 2^5 |\Lambda_{F F F T}|^2 \eta_F^3 \eta_T \frac{Q_F Q_T}{\hbar \omega_T^2 \omega_F}
    P_T
    \text{sinc}^2\left( \frac{{\Delta \kappa\cal{L}}}{2}\right),
    \label{eq:spont_rateA}
\end{align}
where $P_T$ is the pump power in the input channel, and we introduce the escape efficiency $\eta_J = Q_J/Q_{J,C}$, with $Q_{J,C}$ being the quality factor determined solely by the coupling $\gamma_J$ between the channel and the input/output channel waveguide. {Here we identify an important qualitative feature of the SpTOPDC efficiency in a resonant system: The lower scaling with the ring quality factors arises due to the lower scaling with pump power, compared to SFWM and SPDC.}

{For simplicity,} we consider the particular non-degenerate case of $G_1=G_2 \equiv G, G_3 \equiv S$ (see Fig \ref{fig:configuration}b)\footnotemark[2]. For SpTOPDC in this configuration, we have 
\footnotetext[2]{{Configurations where $G_1\neq G_2$ can be analysed in a similar way.}}
\begin{align}
    H_{\text{non-deg}}^{I}(t) =& 
    \nonumber - 3\hbar \Lambda_{G G S P} \mathcal{L}^2 \frac{1}{4\pi^2} \text{sinc}\left( \frac{\Delta \kappa\mathcal{L}}{2}\right) \int dk_1 ... dk_4 F_{G+}^*({k_1}) \\
    \nonumber
    &\times  F_{G+}^*({k_2})F_{S+}^*({k_3})F_{P-}({k_4}) e^{-i(\omega_{k_4P} - \omega_{k_3S} - \omega_{k_2G} - \omega_{k_1G})t}\\
    &\times  a^{\dagger}_G(k_1) a^{\dagger}_G(k_2) 
    a^{\dagger}_S(k_3)\alpha_P(k_4) + H.c.,
    \label{eq:H_NL_spontTG_B}
\end{align}
where $S$ labels the seed mode, $\Delta \kappa = \kappa_{P}-\kappa_{S}-2\kappa_{G}$, and $\Lambda_{G G S P}$ is the nonlinear coupling rate, of the same form as (\ref{eq:lambda}). The corresponding spontaneous triplet generation rate is
\begin{align}
    R_{G G S}^\text{spon} &=9\times2^5
    |\Lambda_{G G S P}|^2
    \frac{\eta_G^2 \eta_S \eta_P\:Q_G Q_S Q_P}{\hbar \omega_P^2 (2 Q_S \omega_G + Q_G \omega_S)}
    P_P
    \text{sinc}^2\left( \frac{\Delta \kappa\mathcal{L}}{2}\right).
    \label{eq:spont_rateB}
\end{align}

We can now turn to the stimulation of this non-degenerate process by a seed beam in mode $S$. Taking (\ref{eq:H_NL_spontTG_B}) and treating the seed field classically by taking $a^{\dagger}_S(k_3) \rightarrow \alpha^*_S(k_3)$, we find the rate of pairs generated in mode $G$ to be
\begin{align}
    R_{G G (S)}^\text{stim} &=9\times 2^6
    |\Lambda_{G G S P}|^2
    \eta_G^2 \eta_S \eta_P
    \frac{Q_G Q_S Q_P}{\hbar^2\omega_P^2 \omega_G \omega_S^2 }
    P_P P_S
    \text{sinc}^2\left( \frac{\Delta \kappa\mathcal{L}}{2}\right),
    \nonumber \\
    &=R_{G G S}^\text{spon}\frac{P_S}{P_\text{vac}}. \label{eq:ratio}
\end{align}
Here we have introduced an effective vacuum power $P_\text{vac}={\hbar\omega_S}\left(2/{\overline{\Gamma}_S}+1/{\overline{\Gamma}_G}\right)^{-1},$
{which can be interpreted as the power associated with the vacuum fluctuations that drive the spontaneous process \cite{liscidini_stimulated_2013}. From Eq. (\ref{eq:ratio}) one can see that 
the improvement of the generation rate in the stimulated regime is equal to $P_S/P_\text{vac}$. The efficiencies of SFWM and SPDC, and their stimulated counterparts, have been derived in similar terms; distinct vacuum powers can be identified in each case, {and the theoretical efficiencies of StTOPDC, SFWM, and SPDC for particular systems can be compared by referring to this earlier work \cite{how_does_it_scale}.}}



To provide a realistic estimate of resonant TOPDC rates, we use Eqs. (\ref{eq:spont_rateA}), (\ref{eq:spont_rateB}), and (\ref{eq:ratio}) with experimental parameters demonstrated in a {composite AlN-Si$_3$N$_4$} microring system engineered for third harmonic generation (THG) \cite{Surya:18}. From the THG efficiency quoted in \cite{Surya:18}, we can infer a nonlinear coupling rate of $\Lambda_{THG} \approx 0.99 s^{-1}$. It should be noted that phase matching was achieved with the pump field in the fundamental spatial mode, and the generated third harmonic field in a higher order mode. By taking $\Lambda_{F F F T} = \Lambda_{THG}$ to estimate the triplet generation efficiency, we assume the same modes are used in our reverse process. That is, we assume that a higher order spatial mode of the driving field at $\omega_T$ could be injected into the microring system. {While this is challenging, we adopt this scenario for the purpose of using an existing system to estimate the generation rates that may soon be accessible. So far, there have been no efforts toward engineering microring systems for StTOPDC, and it is likely that more practical systems could be envisaged.}

For degenerate SpTOPDC, we take the observed quality factors $Q_F = 4\times10^5 $ and $Q_T = 6.4\times10^4$ \cite{Surya:18}, we assume critical coupling ($\eta=0.5$) for all modes, and from the design of the structure we have $\Delta \kappa = 0$. We then find the rate of triplets at the system's output to be $R_{F F F}^\text{spon}/P_T = 0.03 s^{-1}W^{-1}$. This low generation rate suggests that SpTOPDC will remain experimentally impractical in the near term, unless significant improvements can be made to system parameters such as quality factors and mode overlap. Turning to StTOPDC, we take $\omega_S \approx \omega_G \approx \omega_F$, so $Q_S \approx Q_G \approx Q_F$, and $\Lambda_{G G S P} \approx \Lambda_{F F F T}$. With this, we predict $R_{G G (S)}^\text{stim}/P_P P_S = 1.4 \times 10^9 s^{-1} W^{-2}$ {in the AlN-Si$_3$N$_4$ system. For a 100 mW pump and a 10 mW seed, both being within the range of input powers studied in \cite{Surya:18}, we predict $1.4\times10^6$ pairs per second at the system's output.}

We now consider the properties of the photons that could be generated in these non-degenerate TOPDC processes. {In the following, we assume a pulsed pump, and for StTOPDC we take a CW seed; however, more general excitation scenarios with arbitrary pump and seed pulses could be modelled using the same approach}. To first order in the nonlinear interaction,
\begin{align}
    \ket{\Psi} \approx \ket{vac} -\frac{i}{\hbar}\int^{\infty}_{-\infty} dt' H_{NL}^{I}(t') \ket{vac},
    \label{eq:firstorder}
\end{align}
where we take the initial and final times to $\pm \infty$ for convenience.

In the case of non-degenerate SpTOPDC, putting the interaction Hamiltonian (\ref{eq:H_NL_spontTG_B}) into (\ref{eq:firstorder}) yields a ket with the form 
\begin{align}
    \ket{\Psi_{III}} &\approx  \ket{vac} + \beta \ket{III},\\ \nonumber
    \ket{III} &= \frac{1}{\sqrt{2}} \int d{k}_1 d{k}_2 d{k}_3 \phi({k}_1,{k}_2,{k}_3) a^{\dagger}_G(k_1) a^{\dagger}_G(k_2) a^{\dagger}_S(k_3) \ket{vac}, 
\end{align}
where  $|\beta|^2$ is the probability of generating a photon triplet per pump pulse, and 

\begin{align}
    \phi&(k_1,k_2,k_3) = \mathcal{N} F_{P-}\left(\frac{v_G}{v_P}(k_1+k_2) + \frac{v_S}{v_P}k_3 + \frac{1}{v_P}\Upsilon\right) \label{eq:triphoton} \\ \nonumber
    \times & F^*_{G+}(k_1) F^*_{G+}(k_2) F^*_{S+}(k_3) \alpha_P\left(\frac{v_G}{v_P}(k_1+k_2) + \frac{v_S}{v_P}k_3 + \frac{1}{v_P}\Upsilon\right)
\end{align}
is the triphoton wavefunction with the normalization factor $\mathcal{N}$, where we have introduced $\Upsilon = v_P K_P - v_S K_{S} - 2 v_G K_{G}$.

Next we turn to StTOPDC, where we take Eq. (\ref{eq:H_NL_spontTG_B}) with $a^{\dagger}_S(k_3) \rightarrow \alpha^*_S(k_3)$. In doing this, we effectively trace over the mode $S$, so we now identify a biphoton state rather than a triphoton state. {The Hamiltonian is sensitive to the spectral profile of the seed field, so the properties of the pairs depend on the form of $\alpha^*_S(k_3)$; here we deal with the simple case of a CW seed at $k_S$.} We have
\begin{align}
    \ket{\Psi_{II}} &\approx  \ket{vac} + \beta' \ket{II},\\ \nonumber
    \ket{II} &= \frac{1}{\sqrt{2}} \int dk_1 dk_2 \phi(k_1,k_2) a^{\dagger}_G(k_1) a^{\dagger}_G(k_2) \ket{vac},
\end{align}
where  $|\beta'|^2$ is the probability of generating a photon pair per pump pulse. The biphoton wave function is given by

\begin{align}
    \phi(k_1,k_2) &= \nonumber \mathcal{N'}  F_{P-}\left(\frac{v_G}{v_P}(k_1+k_2) + \frac{v_S}{v_P}  k_S + \frac{1}{v_P}\Upsilon\right) F_{G+}^*(k_1) \\ &\times F_{G+}^*(k_2)  \alpha_T\left(\frac{v_G}{v_P}(k_1+k_2) +  \frac{v_S}{v_P}  k_S + \frac{1}{v_P}\Upsilon \right) 
    \label{eq:biphoton},
\end{align}
where $\mathcal{N'}$ is again a normalization constant. 


Comparing Eqs. (\ref{eq:triphoton}) and (\ref{eq:biphoton}), we see that the SpTOPDC triphoton wave function at $k_3 = k_S$ is proportional to the biphoton wave function arising from the corresponding StTOPDC process, seeded by a CW field at $k_S$. From this follows the possibility of using stimulated emission tomography (SET) to characterize the state generated by SpTOPDC \cite{dominguez-serna_third-order_2020}. By characterizing the spectral density of the stimulated $\phi(k_1,k_2)$ for various values of $k_S$, one could build up the joint spectral density of the spontaneous triphoton state, slice by slice. {Recognizing this, one can also see that changing the frequency of the CW seed also changes the spectral properties of the pairs in a nontrivial way, since the dependence of the full triphoton wave function on $k_3$ is nontrivial; this comprises an example of the tunability of the photon pairs from StTOPDC.}

Eq. (\ref{eq:biphoton}) also provides insight into the characteristics of the photon pairs generated by StTOPDC. Its form indicates that an approximately separable biphoton wave function could be generated with the appropriate choice of $\alpha_P(k)$, provided that the generated mode's resonance is much narrower than the resonance at the pump mode. {This latter condition may be naturally satisfied in StTOPDC; the wide range of frequencies and the possible use of higher-order modes could allow for the linewidth at the pump resonance to be significantly different than the linewidths at the generation resonances. Indeed, the quality factors observed in the sample AlN-Si$_3$N$_4$ system satisfy the requirements for generating separable photon pairs; the resonant linewidths at $\omega_G$ and $\omega_P$ are mismatched, with the resonance at $\omega_G$ being the sharper one} \cite{vernonOL, Surya:18}. 

\begin{figure}[h]
    \centering
    \includegraphics[width=0.37\textwidth]{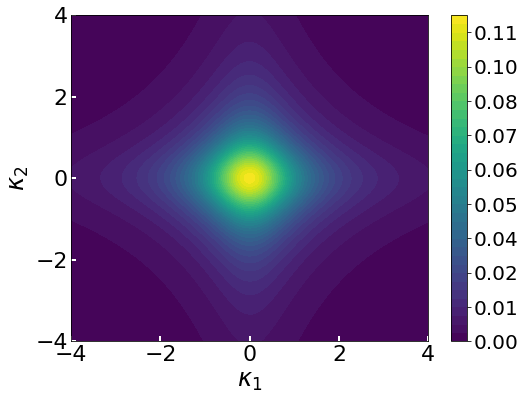}
    \caption{{The normalized JSI describing pairs generated by StTOPDC, plotted in terms of dimensionless variables $\kappa_{1(2)} = v_G(k_{1(2)}-K_G)/\overline{\Gamma}_G$.  The Schmidt number is 1.0016.}}
    \label{fig:JSIs}
\end{figure}

{In Fig. \ref{fig:JSIs} we plot the JSI for StTOPDC using the AlN-Si$_3$N$_4$ parameters quoted in our rate estimate above, and a Gaussian pump pulse with a FWHM of $10$ ps \footnotemark[3]. {The low Schmidt number is attributed to the mismatched linewidths at $\omega_G$ and $\omega_P$, and to the pump duration, which is much shorter than the photon dwelling time in the resonator \cite{Surya:18, vernonOL}}.} {This corroborates the expectation that StTOPDC may be naturally suitable for the generation of unentangled photon pairs; achieving a comparable degree of separability is typically more challenging in SFWM, where carefully engineered microring systems or pumping schemes are required\cite{vernonOL, Christensen}.}
\footnotetext[3]{{Here we focus on a single JSI peak, since we have considered only the generation of photons in a single resonance. In general, multiple JSI peaks will exist, corresponding to the different sets of resonances in which pairs could be generated. The separation of the peaks in frequency space would be set by the ring's free spectral range, which is much larger than the resonance linewidths.}}

{We have presented a simple first treatment of resonant StTOPDC; there is a wealth of extensions to investigate, particularly with more complex pumping schemes and photonic systems. Even using a system that is not optimized for StTOPDC, our rate estimates suggest that this process will be viable in integrated photonic systems within the near future. Further theoretical and experimental work on StTOPDC should reveal qualitatively new aspects of quantum nonlinear optics.}\\




\textbf{Funding.} M.B. acknowledges support from the University of Toronto Faculty of Arts \& Science Top Doctoral Fellowship. J.E.S. and M.B. acknowledge support from the Natural Sciences and Engineering Research Council of Canada.  M.L. acknowledges support by Ministero dell’Istruzione, dell’ Università e della Ricerca (Dipartimenti di Eccellenza Program (2018–2022)).

\textbf{Data Availability.} Data underlying the results presented in this paper are not publicly available at this time but may be obtained from the authors upon reasonable request.

\textbf{Disclosure.} The authors declare no conflicts of interest. 

\bibliography{refs}

\bibliographyfullrefs{refs}

\end{document}